\documentclass[pss,fleqn,embeddedheads]{w-art}
\usepackage{times}
\usepackage{w-thm}
\usepackage{amsmath,amssymb}

\newcommand{\reff}[1]{Fig.\ \ref{fig:#1}}
\usepackage[]{graphicx}
\begin{document}
\renewcommand{\copyrightyear}{2005}
\DOIsuffix{theDOIsuffix}
\Volume{XX}
\Issue{x}
\Month{XX}
\Year{2005}
\Receiveddate{19 July 2005}
\Reviseddate{}
\Accepteddate{}
\Dateposted{}



\title[Orbital-selective Mott transitions]{Orbital-selective Mott transitions in the 2-band $\boldsymbol{J_z}$-model:\\
a high-precision quantum Monte Carlo study}


\author[P.\ G.\ J. van Dongen]{P.\ G.\ J. van Dongen\footnote{Corresponding
     author: e-mail: {\sf Peter.vanDongen@uni-mainz.de}, Phone: +49\,6131\,39-25609
     Fax:  +49\,6131\,39-20954}} \address{Institute of Physics, Johannes Gutenberg University, 55099 Mainz, Germany}
\author[C. Knecht]{C. Knecht}
\author[N. Bl\"umer]{N. Bl\"umer}
\begin{abstract}
  Using high-precision quantum Monte Carlo (QMC) simulations within
  the framework of dynamical mean field theory (DMFT), we show that
  the anisotropic degenerate two-orbital Hubbard model contains two
  consecutive orbital-selective Mott transitions (OSMTs) even in the
  absence of spin-flip terms and pair-hopping processes. In order to
  reveal the second transition
  we carefully analyze the low-frequency part of the self-energy and
  the local spectral functions. This paper extends our previous work
  to lower temperatures. We discuss the nature -- in particular the
  order -- of both Mott transitions and list various possible
  extensions.
\end{abstract}
\maketitle                   





  \section{Introduction}
  Recently, the experimental discovery
  \cite{Nakatsuji00ab} of two consecutive
  orbital-selective metal-insula\-tor transitions in the effective
  3-band system Ca$_{2-x}$Sr$_{x}$RuO$_4$ triggered a cascade of
  papers also in the theoretical literature. The great interest in
  these ``orbital-selective Mott transitions'' can easily be
  understood. After initial LDA and DMFT \cite{Anisimov02}, 
  band structure \cite{Fang01,Fang04}, and
  strong-coupling \cite{Sigrist04} calculations, it became clear that
  the occurence of OSMTs in Ca$_{2-x}$Sr$_{x}$RuO$_4$ is a
  non-perturbative correlation phenomenon, which generalizes the
  famous concept of the classical ``Mott transition'' \cite{Gebhard97}
  to multi-band systems with inequivalent bands.  The
  orbital-selective nature of the transitions also raises fundamental
  questions, such as, e.g., whether phase transitions in two
  subsystems merge or remain distinct, depending on the way these two
  subsystems are being coupled.

  The microscopic model used in the literature to describe the OSMT
  phenomenon theoretically is the anisotropic degenerate two-orbital
  Hubbard model with
  \cite{Koga04a,Koga04b,Ferrero05,deMedici05,Arita05,Koga05} or
  without \cite{Liebsch03ab,Liebsch04,Knecht05a,Knecht05b} spin-flip
  terms and pair-hopping in the Hamiltonian. For the model
  \emph{without} these additional terms (the ``$J_z$-model'') it was
  first believed \cite{Liebsch03ab,Liebsch04} that only
  \emph{one} transition exists at low temperatures, i.e., that the
  transitions which occur in the separate orbitals merge upon unison.
  Since \emph{two} transitions clearly exist for the model \emph{with}
  spin-flips and pair-hopping (the ``$J$-model'') it was believed
  \cite{Koga04b} that these additional terms cause the two transitions
  of the subsystems to remain distinct and, hence, make a fundamental
  difference. Here we show, extending previous work in \cite{Knecht05a}, that the
  $J_z$-model does, in fact, also contain two distinct OSMTs. In this
  manner we establish the $J_z$-model as a \emph{minimal model} for
  the description of the OSMT phenomenon.

  The Hamiltonian of the $J_z$-model,
\begin{equation*}
  H=-\sum_{\langle ij\rangle m\sigma}  t^{\phantom{\dagger}}_m c^{\dag}_{im\sigma}
  c^{\phantom{\dagger}}_{jm\sigma}\,+\,U\sum_{im}n_{im\uparrow} n_{im\downarrow}
    +\sum\nolimits_{i\sigma\sigma'}(U'-\delta^{\phantom{\dagger}}_{\sigma \sigma'} J_z)\,n^{\phantom{\dagger}}_{i1\sigma}
  n^{\phantom{\dagger}}_{i2\sigma'}\quad,
  \end{equation*}
  describes hopping between nearest-neighbor sites $i,j$ with
  amplitude $t_m$ for orbital $m\!\in\!\{1,2\}$, \emph{intra}- and
  \emph{inter}\-orbital Coulomb repulsion parametrized by $U$ and
  $U'$, respectively, and Ising-type Hund's exchange coupling $J_z$;
  $n^{\phantom{\dagger}}_{im\sigma} =c^{\dag}_{im\sigma}
  c^{\phantom{\dagger}}_{im\sigma}$ for spin
  $\sigma\in\{\uparrow,\downarrow\}$. As usual, we consider this model
  with $U'=U-2J_z$. Specifically, our DMFT (QMC) results for this model,
  to be presented below, were calculated using semi-elliptic densities
  of states with full bandwidths $W_1=2$, $W_2=4$, for the ``narrow'' and
  ``wide'' band, respectively, and interaction parameters $J_z=U/4,\; U'=U/2$.
  
  It is important to emphasize that high-precision in the QMC
  simulations is \emph{essential} for the investigation of OSMTs,
  since the signals of in particular the second transition are rather
  subtle and can easily be missed. In \cite{Knecht05b} we showed that
  our QMC results for, e.g., quasi-particle weights, have relative
  errors of ${\cal O}(10^{-2})$ even near the critical interactions
  $U_{c1}\simeq 2.0$ and $U_{c2}\simeq 2.5$ for the first and
  second transition, respectively. This is an improvement of up to two
  orders of magnitude compared to previous QMC results
  \cite{Liebsch03ab,Liebsch04}. The high precision in our
  simulations is (at least in part) due to the reduction of the 
  discretization error achieved by complementing QMC data with a 
  high-frequency expansion of the self-energy \cite{Knecht02,Bluemer04}.

  \vspace{-1em}
  \section{Results}
  In the following, we present QMC results for the
  $J_z$-model with band widths and interaction parameters as stated before. We briefly discuss one of the primary
  criteria used in Ref.\ \cite{Knecht05a} for detecting the two OSMTs in \reff{wImS}
  (cf.\ also inset of \reff{spectra}) before turning to new results.

  As shown in \reff{wImS}, a low-frequency analysis of the self-energies
  \begin{figure}
  \includegraphics[width=0.49\columnwidth,clip=true]{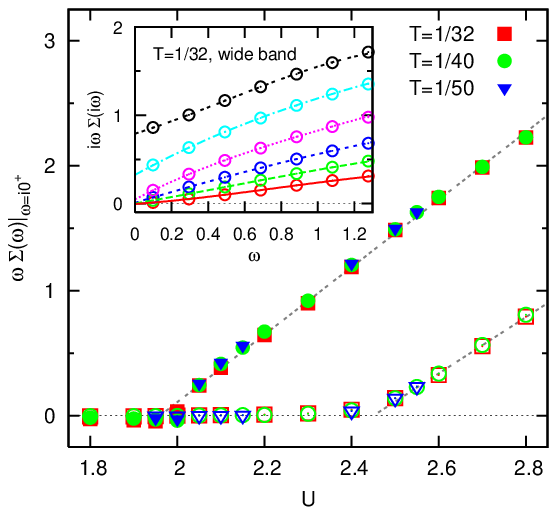}
  \hfill
  \includegraphics[width=0.49\columnwidth,clip=true]{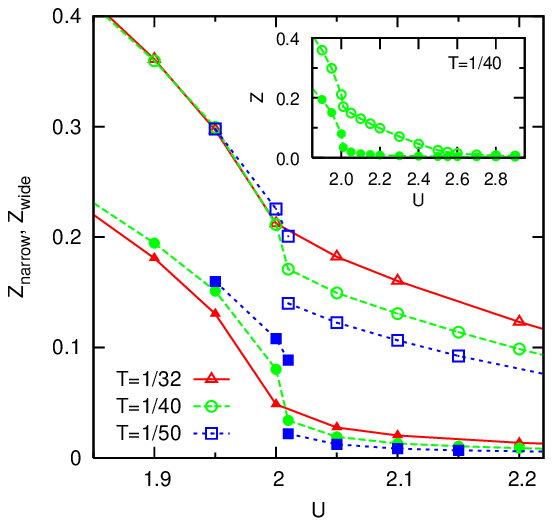}

  \vspace{-0.5em}
  \begin{minipage}[t]{0.48\textwidth}
  \caption{Low-frequency analysis of the self-energy. Main panel:
    weight of singularity at $\omega\!=\!0$ for narrow/wide band
    (filled/open symbols) as extracted from polynomial fits to the
    product $\omega\Sigma(\omega)$ [shown in inset for $T\!=\!1/32$
   and $U=2.8,2.6,2.4,2.2,2.0,1.8$ (top to bottom)].\label{fig:wImS}}
  \end{minipage}\hfill
  \begin{minipage}[t]{0.48\textwidth}
  \caption{Quasiparticle weight $Z\!=\!m/m^*$. Inset: $Z$ for wide/narrow
    band (open/filled symbols) across OSMTs at $U_{c1}\!\approx\!2.0$
    and $U_{c2}\!\approx\!2.5$ for $T\!=\!1/40$. Main panel: detailed
    study of first transition at $U_{c1}$ reveals hysteresis (i.e. a
    $1^{\text{st}}$ order transition) only for $T<T^*\approx
    0.02$.\label{fig:Z}}
  \end{minipage}
  \end{figure}
  associated with both bands clearly reveals the orbital-selective
  character of the first transition at $U_{c1}\approx\! 2.0$
  (associated with the narrow band) and the existence of a second
  transition (associated with the wide band) at $U_{c2}\approx\! 2.5$:
  for $U\!\approx\! U_{c1}$, only the narrow band develops a
  singularity the weight of which (filled symbols) increases
  approximately linearly with $U$.  In contrast, the corresponding
  weight for the narrow band remains zero (numerically) up to
  $U\approx U_{c2}$; above this point, again, a linear increase with
  $U$ is observed.  Note that temperature dependencies are hardly
  visible; however, the kink at $U_{c2}$ becomes slightly sharper with
  decreasing $T$.  The extraction of the weights shown in the main
  panel from fits to products $\omega \Sigma(\omega)$ is illustrated
  in the inset to \reff{wImS}.
  
  Metal-insulator transitions may also be identified from discrete
  estimates of the quasiparticle weight $Z\approx [1-
  \text{Im}\,\Sigma(i\pi T)/(\pi T)]^{-1}$ (even though this
  observable is somewhat problematic \cite{Knecht05a,Knecht05b}):
  kinks in $Z$, most markedly for the narrow band, reveal a transition
  at $U_{c1}\approx\! 2.0$ in the inset of \reff{Z}; a second
  transition for $U_{c2}\approx\! 2.5$ would become apparent at higher
  resolution. The main panel of \reff{Z} shows a close-up of $Z$ near
  $U_{c1}$ for a range of temperatures. Evidently, the observables
  vary continuously with $U$ for $T=1/32$ and $T=1/40$; only for
  $T=1/50=0.02$, we find first traces of stable coexistence, signaling
  a first-order transition. Thus, we estimate $T^*_{c1}\approx 0.02$
  which agrees with recent independent exact diagonalization (ED)
  results \cite{Liebsch05a}, but is much lower than earlier QMC
  and ED estimates (of $T^*_{c1}\approx 0.038$ \cite{Liebsch04} and
  $T^*_{c1}\gtrsim 0.03$ \cite{Liebsch05aV1}, respectively).

  Valuable insight into the nature of phases and transitions can be
  gained from the spectral functions, shown for the narrow and wide band
  in the left and right hand panel of \reff{spectra}, respectively.
  \begin{figure}
  \includegraphics[width=0.49\columnwidth,clip=true]{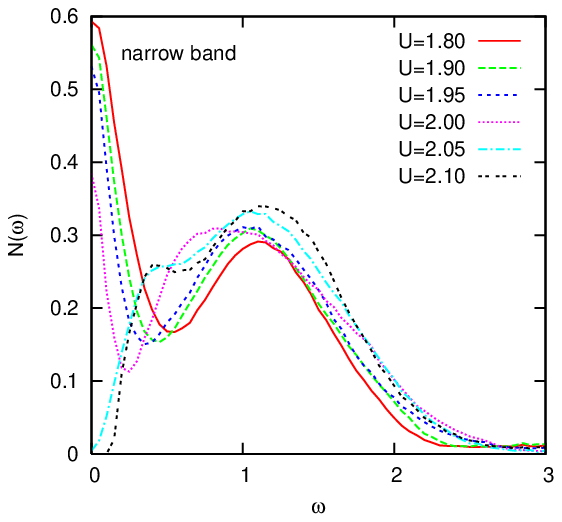}
  \hfill
  \includegraphics[width=0.49\columnwidth,clip=true]{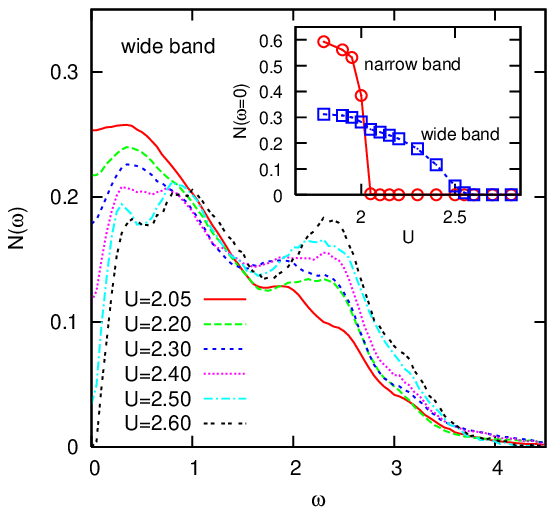}
  \caption{Left panel: Narrow-band spectral function for $T\!=\!1/40$ across
    first OSMT at $U_{c1}\!\approx\! 2.0$. Right panel: Wide-band
    spectral function (for $T\!=\!1/40$) near $U_{c2}\!\approx\! 2.5$.
    Inset: Values of spectral functions at Fermi energy.\label{fig:spectra}}
  \end{figure}
  The narrow-band spectra show typical Fermi-liquid behavior: the
  quasiparticle peak becomes narrow before it decays and, finally, a
  gap appears for $U\gtrsim 2.05$. In contrast, the wide-band
  quasiparticle peak remains wide, but develops a dip near the Fermi
  energy before a full gap appears at $U\approx 2.6$. The density of
  states at the Fermi energy, shown in the inset of \reff{spectra},
  further illustrates the OSMT scenario: At $U_{c1}\approx 2.0$ the
  narrow band undergoes a metal-insulator transition while the wide
  band is hardly affected; the latter becomes insulating only above
  $U_{c2}\approx 2.5$.


  \vspace{-1em}
  \section{Summary and Discussion}
  We have shown that, contrary to previous statements in the
  literature, the two-band Hubbard model with distinct band widths
  $W_2=2 W_1$ and interaction parameters $U,U'\!=\!U/2,J_z\!=\!U/4$
  does in fact describe the occurrence of orbital-selective Mott
  transitions, as seen experimentally in the
  Ca$_{2-x}$Sr$_{x}$RuO$_4$-system. This ``$J_z$-model'' can therefore
  be considered as a \emph{minimal model} for the theoretical
  description of OSMTs. We also showed that the method used in this
  paper, viz.  high-precision QMC calculations at finite temperatures,
  is well-suited for the study of OSMTs, provided the correct
  high-frequency behavior of the self-energy is implemented with care
  and full convergence in the DMFT cycle is established.
  The observables studied in this paper, revealing two consecutive
  Mott transitions, are the spectral functions, the low-energy
  behavior of the self-energy, and the quasiparticle weights.
  
  The question concerning the \emph{order} of the transitions at
  $U_{c1}\simeq 2.0$ and $U_{c2}\simeq 2.5$ is both interesting and
  important. The results presented here suggest that physical
  quantities, like the quasiparticle weights, display a \emph{jump} at
  sufficiently low temperatures for $U=U_{c1}$, so that the first
  transition is likely of first order below $T^*_{c1}\approx 0.02$.
  In contrast, these observables are \emph{continuous} for the
  $J_z$-model at $U=U_{c2}$, as expected for a second-order
  transition. At present it would be premature, however, to make
  definite statements concerning the second transition, because
  high-precision QMC calculations are at present difficult (since
  computationally costly) in the relevant low-temperature regime and
  also because it is numerically difficult to distinguish a real
  second-order transition at $T>0$ from a narrow cross-over, eradiating
  possibly from a $T=0$ quantum critical point.
  In spite of these numerical
  uncertainties, we wish to point out that the occurrence of a
  first-order transition at $U_{c1}$ and a second-order transition at
  $U_{c2}$ are physically plausible. A first-order
  transition for the narrow band at $U_{c1}$ is exactly what one
  expects on the basis of experience with the Mott transition in the
  single-band Hubbard model \cite{Gebhard97,Rozenberg99,Bluemer02,Bluemer04}. The
  orbital-selective phase between $U_{c1}$ and $U_{c2}$ is characterized by
  itinerant electrons in the wide band, interacting with immobile
  (localized) electrons in the narrow band, which is essentially
  Falicov-Kimball physics. Since the metal-insulator transition in the symmetric spinless
  Falicov-Kimball model is of second order (Hubbard-III-like), 
the same scenario could apply to
  the $J_z$-model at $U_{c2}$. Our results further show
  non-Fermi-liquid behavior of the wide band for $U_{c1}\lesssim U<U_{c2}$,
  which is again to be expected on the basis of the Falicov-Kimball
  analogy. The connection to the Falicov-Kimball model was recently
  also pointed out by Biermann et al. \cite{Biermann05}; for a discussion
  of non-Fermi-liquid physics cf.\ also \cite{Liebsch05a}.

  As an outlook, we discuss several extensions of the present work.
  Clearly, since the transition temperature at $U_{c1}$ appears to be
  rather low $(T^*_{c1}\approx 0.02)$, it is important to obtain
  results at even lower temperatures than were considered in this
  paper. The inclusion of spin flips remains an important problem.
  Since the experimental system Ca$_{2-x}$Sr$_{x}$RuO$_4$ has
  \emph{three} bands, two of which are physically equivalent, it may
  be worthwhile to study the OSMTs in a real 3-band-model. For
  comparison with the experimental situation, additional hybridization
  between the bands needs to be taken into account; first results \cite{deMedici05,Koga05}
  suggest that hybridization influences the OSMT dramatically. Since
  anisotropy is by definition important for the OSMT, it is only
  consistent to take also orbital-dependent interaction strengths into
  account. Since the experimental system is an antiferromagnetic
  insulator for small doping $(x\lesssim 0.2)$, it is important to
  extend the present work to include also magnetic phases. In
  conclusion, the study of OSMTs will doubtlessly remain exciting,
  yielding unexpected important results in the months to come.

\begin{acknowledgement}
 We thank K.~Held, E.~Jeckelmann, and K.~Ueda for stimulating
  discussions and acknowledge support by the NIC J\"ulich and by the DFG
  (Forschergruppe 559, Bl775/1).
\end{acknowledgement}

\end{document}